\documentclass[sigconf, anonymous=False]{acmart}

\usepackage{tabularx, booktabs}

\usepackage{multirow}
\usepackage{float} 
\usepackage{amsmath}
\usepackage[ruled,linesnumbered]{algorithm2e}
\usepackage{graphicx}
\usepackage{subfigure}
\usepackage{epstopdf}
\usepackage{tablefootnote}
\usepackage[flushleft]{threeparttable}
\usepackage{diagbox}
\usepackage{booktabs}
\usepackage{siunitx}
\usepackage{footmisc}
\usepackage{footnote}
\usepackage{enumitem}
\usepackage[normalem]{ulem}
\useunder{\uline}{\ul}{}

\makeatletter
\newcommand{\ssymbol}[1]{^{\@fnsymbol{#1}}}
\makeatother

\makeatletter
\def\hlinew#1{%
  \noalign{\ifnum0=`}\fi\hrule \@height #1 \futurelet
   \reserved@a\@xhline}
\makeatother

\AtBeginDocument{%
  \providecommand\BibTeX{{%
    \normalfont B\kern-0.5em{\scshape i\kern-0.25em b}\kern-0.8em\TeX}}}

\setcopyright{acmcopyright}
\copyrightyear{2023}
\acmYear{2023}
\acmDOI{}
\acmISBN{}
\acmConference[WSDM'23]{The 16th International ACM WSDM Conference on Research and Development in Information Retrieval}{from February 27 to March 3, 2023}{}

\begin{document}

\title{Towards Better Web Search Performance: Pre-training, Fine-tuning and Learning to Rank}

\author{Haitao Li}
\affiliation{DCST, Tsinghua University}
\affiliation{Zhongguancun Laboratory}
\affiliation{Beijing 100084, China}
\email{liht22@mails.tsinghua.edu.cn}

\author{Jia Chen}
\affiliation{DCST, Tsinghua University}
\affiliation{Zhongguancun Laboratory}
\affiliation{Beijing 100084, China}
\email{chenjia0831@gmail.com}

\author{Weihang Su}
\affiliation{DCST, Tsinghua University}
\affiliation{Zhongguancun Laboratory}
\affiliation{Beijing 100084, China}
\email{swh22@mails.tsinghua.edu.cn}

\author{Qingyao Ai}
\affiliation{DCST, Tsinghua University}
\affiliation{Zhongguancun Laboratory}
\affiliation{Beijing 100084, China}
\email{aiqy@tsinghua.edu.cn}

\author{Yiqun Liu}
\authornote{Corresponding author}
\affiliation{DCST, Tsinghua University}
\affiliation{Zhongguancun Laboratory}
\affiliation{Beijing 100084, China}
\email{yiqunliu@tsinghua.edu.cn}

\begin{abstract}
This paper describes the approach of the THUIR team at the WSDM Cup 2023 Pre-training for Web Search task. This task requires the participant to rank the relevant documents for each query. We propose a new data pre-processing method and conduct pre-training and fine-tuning with the processed data. Moreover, we extract statistical, axiomatic, and semantic features to enhance the ranking performance. After the feature extraction, diverse learning-to-rank models are employed to merge those features. The experimental results show the superiority of our proposal. We finally achieve second place in this competition.
\end{abstract}

\ccsdesc[500]{Information systems~Retrieval models and ranking}
\ccsdesc[300]{Information systems~Information retrieval}

\keywords{WSDM Cup 2023, Pre-training, Web Search, Learning to rank}

\maketitle

\section{Introduction}
In recent years, pre-trained language models (PLMs) have been proved effective in capturing rich language knowledge from large-scale corpus, reaching state-of-the-art performance in many downstream natural language processing (NLP) tasks~\cite{devlin2018bert, vaswani2017attention, yang2019xlnet, liu2019roberta, yasunaga2022linkbert}. These pre-trained language models are first trained on a large-scale corpus without supervision and then fine-tuned on labeled data for downstream tasks. An ideal PLM for Information Retrieval (IR) should have the ability to measure the relevance between a query and the candidate document. However, the pre-training tasks in the NLP field such as Masked Language Modeling (MLM) and Next Sentence Prediction (NSP), cannot directly enhance the relevance-matching ability of PLMs. Thus, researchers in the IR field have begun to explore the pre-training methods tailored for Web Search~\cite{ma2021prop, ma2021b, chang2020pre,fan2021pre,chen2022axiomatically}. 

Thanks to the Pre-training for Web Search challenge in WSDM Cup 2023, a real-world long-tail user feedback dataset from Baidu Search is now available for exploring better pre-training methods for web search. The competition's goal is to achieve better ranking performance by leveraging advanced pre-training and fine-tuning methodologies. 

In this paper, we introduce the solution of THUIR team, which achieves second place in the competition (DCG@10 10.04097 on the leaderboard). To be specific, we first conduct an exploratory analysis of the pre-training corpus and filter out a new corpus. Then, we pre-train a 12-layer transformer with MLM and CTR prediction tasks and fine-tune it with the expert annotated data. Finally, we extracted 20 features for each query-document pair. Diverse learning-to-rank methods are employed to aggregate these features into the final relevance score. 

This paper is organized as follows: Section 2 introduces the pre-training and fine-tuning dataset. In Section 3, details of our method is elaborated. Then, the experimental setting and results are introduced in Section 4. Finally, we conclude our work and discuss the future direction in Section 5.

\section{Dataset}
In this section, we focus on the datasets provided by the competition.
\subsection{Large Scale Web Search Session Data}
The large-scale web search session data is real long-tail user feedback data provided by Baidu~\cite{zou2022large}. This data contains 2000 files in .gz format and each file is about 500M. This session data only provides the pre-processed tokens of the query and document.

To construct our pre-trained corpus, we first encode queries in both session data and expert annotation data into vectors with the pre-trained language model. Then, the inner product is employed to calculate the similarity of the two types of queries. If a query on session data has a similarity of less than 0.9 with all the queries annotated by experts, then we will filter it out from the session data. Moreover, we filtered out queries that did not click on any documents. We think these queries lack positive samples and cannot provide sufficient training information for the pre-trained language models.

\subsection{Expert Annotation Dataset}
Expert annotated data contains 7,008 queries and 397,552 query-document pairs. Since the official data has the problem of different queries with the same id, we first mix the original training set and the validation set and then remap the QIDS for all querise and ensure one query only have one unique identifier. After that, we encode all queries into vectors with the pre-trained language model. We calculate the similarity between the above queries and testing queries and select 20\% most similar queries as the new validation set.

Finally, there are 298,537 training instances and 99,015 validation instances, which are named traing\_new and test\_new respectively. Furthermore, we count the occurrence of all tokens and select the token with the top 50 occurrence frequency as the stop words.

\section{Method}
In this section, we present the complete solution of the Pre-training for Web Search task.

\subsection{Pre-training}

As pre-trained language models (PLMs) have shown great effectiveness in IR tasks, we adopt PLMs as our backbone model. To be specific, we used the official pre-training code released by Baidu to train a backbone model. Two tasks are employed in our method: MLM and CTR tasks. The loss function of these two tasks is as follows:

\begin{equation}
    \label{MLM}
\mathcal{L}_{M L M}=-\sum_{\hat{x} \in m(x)} \log p\left(\hat{x} \mid x_{\backslash m(x)}\right)
 \end{equation}

\begin{equation}
    \label{CTR}
\mathcal{L}_{CTR}=-\sum_{c \in C} c \log (\mathrm{s})+(1-\mathrm{c}) \log (1-\mathrm{s})
 \end{equation}
where $x$, $m(x)$, and $x_{\backslash m(x)}$ denote the input sequence, the masked and the rest word sets in x, respectively. Besides, $C$ is the complete click feedback set, $c$ is the click signal on a result, and $s$ is the estimated click probability of this result. We set the mask rate of the MLM task as 40\% and stop the pre-training process if the total loss does not decline for 1k steps.

\subsection{Fine-tuning}

For each query, the candidate documents are labeled with a relevance score $\in \{0, 1, 2, 3, 4\}$. We define the candidate documents with labels 3 and 4 as positive document ${d^+}$ and the candidate documents with labels 0 and 1 as ${d^-}$.

We use the Transformer encoder with a similar structure with BERT as the ranking model. The score of a query document pair is defined as follows,
\begin{equation}
Input = [CLS] query [SEP] title [SEP] content [SEP]
 \end{equation}
 
\begin{equation}
 score(query,doc) = MLP( CLS[BERT(Input)] ) 
 \end{equation}
where CLS is the [CLS] token vector and MLP is a Multilayer Perceptron that projects the CLS vector to a score. We use the Cross Entropy Loss to optimize the re-ranking and retrieval model, which is defined as:
\begin{equation}
Loss= -y \log \hat{y} - (1-y) \log (1- \hat{y})
 \end{equation}
where $y$ is the ground truth and $\hat{y}$ is the predicted probability.

\subsection{Learning to Rank}
Learning-to-rank, a popular machine learning method for ranking, has been widely used in information retrieval and data mining~\cite{suthuir2,yang2022thuir}. In this section, we implement several learning-to-rank methods with diverse features.
\subsubsection{Dataset}
We adopt traing\_new as the training data and the valid\_new as the validation set. There are 298,537 training pairs and 99,015 validation pairs and there is no overlap between them. The relevance labeling is in a 5-level setting ranging from 0 to 4 with increasing relevance.
\begin{table}[t]
\caption{
Final features to train learning to rank models. BM25\_title and BM25\_content are only considered for the title and content of the document respectively. PROX-nonstop denotes the model results considering the stop words.
}
\label{feature}
\begin{tabular}{cl|cl}
\hline
Feature ID & Feature Name    & Feature ID & Feature Name \\ \hline
1          & cross encoder   & 11         & TF           \\
2          & query\_length   & 12         & IDF          \\
3          & title\_length   & 13         & PROX-1       \\
4          & content\_length & 14         & PROX-2       \\
5          & query\_freq     & 15         & PROX-3       \\
6          & BM25            & 16         & PROX-4       \\
7          & QLD             & 17         & PROX-1-nonstop  \\
8          & BM25\_title     & 18         & PROX-2-nonstop  \\
9          & BM25\_content   & 19         & PROX-3-nonstop  \\
10         & TF-IDF          & 20         & PROX-4-nonstop  \\ \hline
\end{tabular}
\end{table}
\subsubsection{Feature Extraction}
Feature engineering plays an essential role in learning to rank. Except for taking the scores of the above pre-trained language models as features, we also construct statistical features, axiomatic features~\cite{chen2022axiomatically}, and semantic features to achieve better ranking performance. All the extracted features are shown as follows:

\begin{itemize}[leftmargin=*]
    \item \textbf{Statistical Features}
    
    \begin{itemize}
        \item[-] \textbf{query\_length} is the length of the query token.
        \item[-] \textbf{title\_length} is the length of the document title token.
        \item[-] \textbf{content\_length} is the length of the document content token.
        \item[-] \textbf{query\_freq} is the monthly search frequency of query. The queries are descendingly split into 10 buckets according to their monthly search frequency.
        \item[-] \textbf{TF-IDF}~\cite{ramos2003using} is a classical retrieval model based on bag-of-words. TF-IDF is the combination of term frequency (TF) and inverse document frequency (IDF). Their equations are shown as follows:
        \begin{equation}
	TF(t_{i,j}) = \dfrac{n_{i,j}}{\sum_{k}n_{k,j}}
	\label{eq:TF calculation}
        \end{equation}
        
        \begin{equation}
	IDF(t_i) = \log \dfrac{|D|}{|D_i+1|}
	\label{eq:IDF calculation}
        \end{equation}

        \begin{equation}
	TF-IDF = TF \times IDF
	\label{eq:TFIDF calculation}
        \end{equation}
        where $n_{i,j}$ denotes the number of words $t_i$ in the document $d_j$, $D$ represents the total number of documents in the corpus and $D_i$ represents the number of documents containing the word $t_i$. We extracted the TF, IDF, TF-IDF of the title, content, and title + content respectively.

        \item[-] \textbf{BM25}~\cite{robertson2009probabilistic} is a highly effective sparse retriever based on exact word matching. The calculation
        formula of BM25 is shown in Eq ~\ref{eq:BM25 calculation}.
        \begin{equation}
	BM25(d, q) = \sum_{i = 1}^M \dfrac{IDF(t_i) \cdot TF(t_i, d)       \cdot (k_1+1)}{TF(t_i, d) + k_1 \cdot \left(1-b+b \cdot           \dfrac{len(d)}{avgdl}\right)}
	\label{eq:BM25 calculation}
    \end{equation}
    where $k_1$, $b$ are hyperparameters. Grid search is employed to determine the optimal hyperparameters i.e. $k_1$ = 1.6 and $b$ = 0.87. Similar to TF-IDF, we extracted three scores of BM25: title, content and title+content.
    \item[-] \textbf{QLD}~\cite{zhai2008statistical} is another representative traditional retrieval model based on Dirichlet smoothing. The equation for QLD is shown in Eq ~\ref{eq:language model calculation}. We only extracted the QLD score of title+content.
    
    \begin{equation}
    	\log p(q|d) = \sum_{i: c(q_i; d)>0} \log \dfrac{p_s(q_i|d)}{\alpha_d p(q_i|\mathcal{C})} + n \log \alpha_d +\sum_i \log p(q_i|\mathcal{C})
    	\label{eq:language model calculation}
    \end{equation}

    \end{itemize}  

    \item \textbf{Axiomatic Features} 
    
   \begin{itemize}
        
        \item[-] \textbf{PROX\_1}~\cite{hagen2016axiomatic}: Averaged proximity score of query terms in the candidate document.
        \item[-] \textbf{PROX\_2}~\cite{hagen2016axiomatic}: First appearance position of query terms appearing in the candidate document.
        \item[-] \textbf{PROX\_3}~\cite{hagen2016axiomatic}: Number of query term pairs appearing in the candidate document within a distance of 5.

        \item[-] \textbf{PROX\_4}~\cite{hagen2016axiomatic}: Number of query term pairs appearing in the candidate document within a distance of 10. 
        \item[-] \textbf{STM\_1}~\cite{fang2006semantic}: The STM-1 axiom prefers to select queries that have a high semantic similarity to the document.
        \item[-] \textbf{STM\_2}~\cite{fang2006semantic}: The STM-2 axiom is used to distinguish two queries if they have a close similarity to the same document.
        \item[-] \textbf{REG}~\cite{wu2012relation}: The REG axiom tends to select those queries where the most diverse terms appear more often in the document.
    \end{itemize}
    
    \item \textbf{Semantic Features}:
    
    \begin{itemize}

        \item[-] \textbf{Dense Embedding}:
        The Transformer encoders (e.g., BERT, RoBERTa) output a contextualized vector for each input token and we select the CLS vector as the semantic embedding of a document. Then, we use dot prodouct to get the score of the query and document as the feature

    \end{itemize}
\end{itemize}

\subsubsection{Modeling}
We employ four methods of learning to rank: Lightgbm, XGBoost, LambdaMART and Coordinate Ascent. For LambdaMART and Coordinate Ascent, Ranklib ~\footnote{\url{https://github.com/codelibs/ranklib}} package is applied to implement them. For Lightgbm and XGBoost, we reproduce them with their official packages.
 
\section{Experiment}
\subsection{Feature Selection}
We generate tens of features, and the learning ranking models are trained based on these features. However, there are complex homogenization relationships between features and many features are distributed inconsistently between the training and test sets. To make the model more effective, we further perform feature analysis and selection.
To be specific, we perform heuristic feature selection based on ablation experiments. With fixed model hyperparameters and random seeds, we eliminate one or a set of similar features at a time to train the learning-to-rank model. After feature selection, 20 features are reserved to train the final model. The final features are displayed in Table ~\ref{feature}.

\begin{table}[t]
\caption{Overall experimental results of different methods. The best compression method in each column is marked in bold. }
\label{result}
\begin{tabular}{@{}lll@{}}
\toprule
Model               & Vaild Set      & Leaderboard    \\ \midrule
BM25(k1=1.6,b=0.86) & 10.30          & 9.51           \\
TF                  & 9.14           & -              \\
IDF                 & 8.92           & -              \\
TF-IDF              & 9.20           & -              \\
QLD                 & 9.84              & -              \\
Cross encoder       & 10.06          & 8.58           \\
Coordinate Ascent   & 11.04              & -              \\
LambdaMART          & 11.11              & -              \\
XGBoost             & 11.18          & 10.00          \\
LightGBM            & \textbf{11.25} & \textbf{10.04} \\ \bottomrule
\end{tabular}
\end{table}

\subsection{Experimental Results}
The performance comparisons of different methods are shown in Table \ref{result}. DCG@10 is employed to evaluate the performance of ranking systems. We derive the following observations from the experiment results.

\begin{itemize}[leftmargin=*]
    \item Traditional lexical matching models such as BM25, QL, and TF-IDF show promising performance. It can be observed that BM25, which has been carefully tuned, achieves a score of 9.51 on the leaderboard, which exceeds most teams in the competition.
    \item Despite the extensive training data, the performance of the cross encoder is not satisfactory. After careful examination, we find that the official paddle code may exist bugs. However, there is no time to correct this mistake for our team. Even so, the score of the cross encoder still improves the final learning to rank model, which indicates that the cross encoder provides the information on different aspects of the dataset.
    \item Compared to a single model, learning to rank models achieve better performance, which indicates that fusing different information from different combinations of datasets is beneficial for ranking. Consequently, LightGBM achieves the highest score and yields our final submission.
\end{itemize}

\section{Conclusion}
This paper describes our solution for WSDM Cup 2023 Pre-training for Web Search task. We extract statistical, axiomatic, and semantic features to enhance the ranking performance. We finally achieve second place in this competition. In the future, we will explore more pre-training methods for web search based on search engine logs. 

\bibliographystyle{ACM-Reference-Format}
\bibliography{sample-base.bib}
\end{document}